\documentclass[prd,floatfix,onecolumn,amsmath]{revtex4-1}
\usepackage{graphicx,color,dcolumn,booktabs,bm}
\usepackage{subfigure}
\usepackage{longtable}
\usepackage{amssymb}
\usepackage{indentfirst}
\usepackage{epsfig}
\usepackage{feynmf}   
\usepackage{epstopdf}   
\usepackage{slashed}  
\usepackage{cases}
\usepackage[pdfpages]{xcolor}
\usepackage{pgf}
\definecolor{maroon}{RGB}{139,25,150}
\usepackage{multirow}
\usepackage{physics}
\usepackage{float}
\usepackage[colorlinks, citecolor=magenta,anchorcolor=red,menucolor=blue, linkcolor=blue,filecolor=red,runcolor=blue,urlcolor=magenta,frenchlinks=blue]{hyperref}

\begin{document}

	\preprint{}
	
\title{\color{maroon}{Global Analyses of Generalized Parton Distributions with Diverse PDF Inputs}}

	\author{The MMGPDs Collaboration:\\
Fatemeh Irani$^{1}$}
\email{f.irani@ut.ac.ir}

\author{Muhammad Goharipour$^{2,3}$}
\email{muhammad.goharipour@ipm.ir}
\thanks{Corresponding author}

\author{K.~Azizi$^{1,4,3}$}
\email{kazem.azizi@ut.ac.ir}

\affiliation{
$^{1}$Department of Physics, University of Tehran, North Karegar Avenue, Tehran 14395-547, Iran\\
$^{2}$School of Physics, Institute for Research in Fundamental Sciences (IPM), P.O. Box  19395-5531, Tehran, Iran\\
$^{3}$School of Particles and Accelerators, Institute for Research in Fundamental Sciences (IPM), P.O. Box 19395-5746, Tehran, Iran\\
$^{4}$Department of Physics, Dogus University, Dudullu-\"{U}mraniye, 34775 Istanbul, T\"urkiye }

\date{\today}

\begin{abstract}

This paper investigates the crucial role of parton distribution functions (PDFs) in high-energy physics, particularly their impact on the extraction of generalized parton distributions (GPDs) at zero skewness. To this aim, we perform six global analyses of GPDs using different modern PDF sets (\texttt{NNPDF40}, \texttt{CT18}, and \texttt{MSHT20}) at three specific factorization scales ($ \mu = 2 $, $ 1.3 $, and $ 1 $ GeV) and different perturbative orders. A wide range of elastic electron scattering data is included in the analysis to constrain GPDs in a broad interval in the momentum transfer squared $ t $. The analyses reveal that the best overall description of experimental data is achieved using \texttt{NNPDF40} PDFs at the next-to-leading order (NLO), with moderate sensitivity to the choice of PDF input, especially in the region of low $ t $. The dependence on the perturbative order is relatively mild, indicating stability in the extraction procedure. We also show that different GPD sets become more consistent at larger $ |t| $ values, with down-quark GPDs experiencing greater suppression than up-quark GPDs. We show that the differences among the
GPD parametrizations propagate into measurable quantities such as proton radii, leading to non-negligible variations in the extracted central values and uncertainties. The six extracted GPD sets, available at different scales and perturbative orders, provide valuable resources for future theoretical and phenomenological studies, offering flexibility for researchers in exploring the proton's internal structure.

\end{abstract}

\maketitle

\section{Introduction}\label{sec:one} 

Parton distribution functions (PDFs) are fundamental quantities in quantum chromodynamics (QCD) that describe the momentum distributions of partons (quarks, antiquarks, and gluons) inside hadrons, particularly nucleons~\cite{Lorce:2025aqp}. More precisely, a PDF $ f_i(x,\mu^2) $ represents the number density finding a parton of flavor $ i $ carrying a fraction $ x $ of the parent hadron's momentum at a given energy scale $ \mu^2 $. Actually, PDFs encode the nonperturbative structure of nucleons and are critical inputs for the theoretical predictions of high-energy scattering processes, such as those occurring in deep inelastic scattering (DIS) and hadron colliders like the Large Hadron Collider (LHC). PDFs cannot yet be determined with sufficient precision over the full kinematic range purely from first-principles calculations, although significant progress has been made using lattice QCD approaches. In addition, while various nonperturbative models exist for calculating PDFs theoretically~\cite{Manohar:1983md,Holtmann:1996be,Traini:2013zqa,Gao:2022uhg,Gao:2023lny,Yu:2024qsd,Gao:2026hix,Lin:2025hka}, they are typically determined through global analyses of experimental data, combined with perturbative QCD (pQCD) evolution. For examples of the most up-to-date PDFs, which employ a variety of approaches and incorporate increasingly comprehensive datasets, see Refs.~\cite{Hou:2019efy,Bailey:2020ooq,Cocuzza:2021cbi,Cridge:2021pxm,NNPDF:2021njg,Ball:2022qtp,Guzzi:2022irq,PDF4LHCWorkingGroup:2022cjn,Kassabov:2023hbm,Accardi:2023gyr,Sitiwaldi:2023jjp,Alekhin:2023uqx,Jing:2023isu,Fu:2023rrs,Ablat:2023tiy,Bonvini:2023tnk,McGowan:2022nag,Cridge:2023ryv,NNPDF:2024nan,Barontini:2024dyb,Cooper-Sarkar:2024crx,Cridge:2023ozx,NNPDF:2024djq,Cridge:2024exf,Alekhin:2024bhs,Flett:2024iex,Ablat:2024muy,MSHT:2024tdn,Ablat:2024uvg,Azizi:2024swj,Gao:2025hlm,Ball:2025xgq,Chu:2025jsi,MoralFigueroa:2025cev,Ball:2025xtj,Cruz-Martinez:2026rct,Cocuzza:2026vey}. The precision of PDFs is critical for testing the Standard Model, probing new physics beyond it, and interpreting collider data with high accuracy.

The evolution of PDFs with respect to the energy scale $\mu^2$ is one of the crucial steps in every QCD global analysis, as experimental data are measured at different energy scales. This evolution is governed by the Dokshitzer-Gribov-Lipatov-Altarelli-Parisi (DGLAP) equations~\cite{Gribov:1972ri,Altarelli:1977zs,Dokshitzer:1977sg}, a set of integro-differential equations derived from pQCD. These equations describe how the PDFs change as the resolution scale $\mu^2$ varies. To be more precise, they govern the effects of parton emissions and splittings, such as $q \to qg$, $g \to q\bar{q}$, and $g \to gg$. The DGLAP equations are formulated as a coupled system, with splitting functions $P_{ij}(z)$. These functions quantify the probability of a parton of type $j$ emitting a parton of type $i$ which carries a fraction $z$ of the parent parton's momentum. Solving these equations allows us to evolve PDFs from an initial scale $\mu_0^2$ to higher scales relevant for high-energy experimental data. It is worth noting in this context that PDFs are parametrized at an initial scale $\mu_0^2$ (typically 1-4 GeV$^2$) and determined through a global $\chi^2$ minimization procedure that incorporates a wide range of experimental data, as mentioned before. The theoretical framework and numerical implementation of DGLAP evolution have been extensively developed over the past decades~\cite{Curci:1980uw,Furmanski:1980cm,Spiesberger:1994dm,Kosower:1997hg,Schoeffel:1998tz,vanNeerven:1999ca,vanNeerven:2000uj,Ratcliffe:2000kp,Weinzierl:2002mv,Vogt:2004ns,Roth:2004ti,Cafarella:2005zj,Cafarella:2008du,Salam:2008qg,Botje:2010ay,Bertone:2013vaa,Carrazza:2014gfa,Bertone:2015cwa,Zarrin:2016kxf,Mottaghizadeh:2017vef,Bertone:2017gds,Goharipour:2018dsa,McGowan:2022nag,Cridge:2023ryv,Candido:2022tld,Lappi:2023lmi,Markovych:2023tpa,Yin:2023dbw,Duwentaster:2023mbk,NNPDF:2024nan,Barontini:2024dyb,Cooper-Sarkar:2024crx,Wang:2024rso,NNPDF:2024dpb,Boroun:2024knv,deFlorian:2025yar,Kniehl:2025ttz,Hampson:2025pvi}. In particular, developments in higher-order calculations and evolution schemes have refined the accuracy of DGLAP evolution. Actually, the implementation of next-to-next-to-leading order (NNLO) and, more recently, next-to-NNLO (N$^3$LO) calculations~\cite{McGowan:2022nag,Cridge:2023ryv,NNPDF:2024nan,Barontini:2024dyb,Cooper-Sarkar:2024crx,Hampson:2025pvi} in evolution codes has been crucial for reducing theoretical uncertainties in modern PDF analyses.

The forward evolution of PDFs from a lower scale $\mu_0^2$ to higher scales $\mu^2$ is now well established through the DGLAP equations. However, the backward evolution, i.e., evolving PDFs from a higher scale back to a lower scale, presents unique challenges~\cite{Botje:2010ay}. 
In fact, backward evolution is formally possible within perturbative QCD; however, it may suffer from numerical instabilities, enhanced sensitivity to higher-order corrections, and potential extrapolation into regions where perturbation theory becomes unreliable. For instance, at higher scales, the contributions from sea quarks and gluons become more significant, making it difficult to disentangle their effects when evolving backward. Additionally, numerical instabilities and ambiguities in the choice of initial conditions at the higher scale can lead to uncertainties in the backward-evolved PDFs. These issues are particularly relevant in global analyses that combine data from multiple experiments at different energy scales, where consistent treatment of evolution in both directions is essential for achieving reliable results. Addressing these challenges requires careful theoretical and computational approaches, including the use of higher-order calculations and advanced evolution schemes to minimize uncertainties.

In global analyses of PDFs, it is common to exclude experimental data with $ \mu^2 $ values below the initial scale $ \mu_0^2 $. This avoids the need for backward evolution during the fitting procedure.  After completing their analyses, phenomenological groups often release grids that allow users to generate the resulting PDFs at arbitrary values of $ x $ and $ \mu^2 $ over a broad range. But, these grids typically don't include $ \mu^2 $ values below $ \mu_0^2 $, corresponding to the region of backward evolution. In the present study, we aim to investigate whether using PDFs at $ \mu^2 $ values below $ \mu_0^2 $ leads to unphysical results. By performing some QCD analyses of the generalized parton distributions (GPDs), we investigate how using PDFs at $ \mu^2 $ values below $ \mu_0^2 $ affect the final results. We present various sets of GPDs at zero skewness that are obtained using different PDF sets with different values of $ \mu^2 $ at both next-to-leading order (NLO) and NNLO. We show that the observed differences in the extracted GPDs-especially in the low-t region-can potentially affect predictions for observables sensitive to GPDs, such as the nucleon electromagnetic radii.

\section{Evolution of PDFs}\label{sec:two}
As mentioned before, PDFs describe the number density of partons (such as a quark or gluon) within a hadron, carrying a specific momentum fraction $ x $, and depend on the scale $ \mu^2 $, which is characteristic of the hard scattering process. PDFs are a fundamental concept in particle physics, particularly within the framework of QCD. In pQCD, the cross section for a hard scattering process or a collision involving nucleons is calculated as the convolution of the partonic cross section (hard scattering coefficients in the case of DIS) with the PDFs of the colliding nucleons. While the $ x $ dependence of the parton densities is nonperturbative, the $ \mu^2 $ dependence can be described within pQCD using the DGLAP evolution equations~\cite{Gribov:1972ri,Altarelli:1977zs,Dokshitzer:1977sg}. 
In fact, DGLAP equations describe how parton distributions evolve as the energy scale $ \mu^2 $ changes,
\begin{equation}\label{dglap}
 \frac{\partial f_i(x, \mu^2)}{\partial \ln{\mu^2}} = \sum_j \int_x^1 \frac{dy}{y} P_{ij}\left(\frac{x}{y}, \alpha_s(\mu^2)\right) f_j(y, \mu^2).
\end{equation}
In the above equation, $ f_i(x, \mu^2) $ is the PDF for a parton of type \(i\) and $ P_{ij}(x/y, \alpha_s(\mu^2)) $ is the splitting function, as discussed earlier.  $ \alpha_s $ is the QCD strong coupling constant~\cite{Gribov:1972ri,Altarelli:1977zs,Dokshitzer:1977sg}.

Note that $ \alpha_s $ evolves with the renormalization scale $ \mu_R^2 $, and its evolution typically starts from an initial value at $ m_Z^2 $ where $ m_Z $ is the $ Z $ boson mass. In analogy, the PDFs evolve with the factorization scale $ \mu_F^2 $, and the initial value for this evolution is determined by the $ x $ dependence of the PDFs at the starting scale $ \mu_0^2 $. When $ \mu_R^2 $ and $ \mu_F^2 $ are set equal, the formalism of QCD evolution becomes relatively straightforward; otherwise, it becomes more complex. Therefore, in this context, all scales $ \mu_R^2 $ and $ \mu_F^2 $ are considered equal to $ \mu^2 $ (i.e., $ \mu_R^2 = \mu_F^2 = \mu^2 $)~\cite{Botje:2010ay}.
Solving the DGLAP equations numerically requires initial conditions for the PDFs at a starting scale $\mu^2_0$, usually determined from experimental data as mentioned before. The scale dependence of $\alpha_s(\mu^2)$ is also required, obtained by numerically integrating its renormalization group equation. The QCD coupling constant plays a key role in common approaches for solving the DGLAP evolution equations~\cite{Salam:2008qg,Bertone:2013vaa,Botje:2010ay}.

Several numerical methods exist for solving the DGLAP evolution equations. Each one has its own benefits and way of being used.
The $ x $-space methods work with PDFs as functions of $x$ directly. These methods solve the integro-differential equations numerically with respect to both $x$ and $\mu^2$. The PDFs are represented on a discrete grid in the variable $x$. The convolution integrals are calculated using numerical methods, which turn the equations into a system of coupled ordinary differential equations (ODEs) for the PDF values at each grid point. 
 These ODEs are solved using standard numerical integrators, such as Runge-Kutta algorithms. Some well-known programs that use $ x $-space methods are \texttt{HOPPET}~\cite{Salam:2008qg}, \texttt{QCDNUM}~\cite{Botje:2010ay}, and \texttt{APFEL}~\cite{Bertone:2013vaa}. 
 These codes are often used within fitting frameworks such as \texttt{xFitter}~\cite{xFitter:2022zjb}, which enables PDF determinations from experimental data.
  
The Mellin space methods ($ N $ space) use the Mellin transform, defined as $\tilde{f}(N) = \int_0^1 dx\, x^{N-1} f(x)$. The main advantage of this approach is that convolutions in $x$ space become simple multiplications in Mellin space~\cite{Kosower:1997hg}. For example, the integral in Eq.~(\ref{dglap}) represents a convolution,
$
(f \otimes g)(x) \equiv \int_x^1 \frac{dz}{z} f(z) g\left(\frac{x}{z}\right),
$
which allows Eq.~(\ref{dglap}) to be rewritten as,
\begin{equation}
\label{conv:dglap}
\frac{\partial f_i}{\partial \ln \mu^2} = \sum_{j=q,\overline{q},g} P_{ij} \otimes f_j\,.
\end{equation}
Applying the Mellin transform to the DGLAP equations converts them into a system of ordinary linear differential equations for the Mellin moments $\tilde{f}_i(N, \mu^2)$,
\begin{equation}
\label{mellin:dglap}
\frac{d \tilde{f}_i(N, \mu^2)}{d \ln \mu^2} = \sum_j \tilde{P}_{ij}(N, \mu^2) \tilde{f}_j(N, \mu^2)\,.
\end{equation}
These ODEs are generally simpler to solve compared to the original integro-differential equations in $x$ space.
After solving for the moments, an inverse Mellin transformation is used to get the PDFs back in $x$ space. An example of a Mellin-space code is \texttt{QCD-PEGASUS}~\cite{Vogt:2004ns}. The analytical solution in Mellin space involves solving these linear ODEs for the moments.
For NNLO calculations, approximations are sometimes necessary because the exact solutions can be too complex to handle~\cite{Vogt:2004ns,deFlorian:2025yar}.

Finally, other analytical methods can sometimes be used, although complete solutions are usually not possible. Some special cases can be solved using series expansions and focus on formulating and solving the DGLAP evolution equations directly for measurable structure functions~\cite{Lappi:2023lmi} or exact formulas between heavy quark thresholds~\cite{Yin:2023dbw}. Analytical techniques such as the Laguerre method~\cite{Furmanski:1981ja,Toldra:2001yz}, Laplace transform technique and Jacobi polynomial approach~\cite{Salajegheh:2018hfs}, and ``brute-force" iterations~\cite{Cabibbo:1978ez} have also been proposed, and various other analytical approaches have been explored in the literature~\cite{Ball:1994kc,Kotikov:1998qt,Mankiewicz:1996sd}.
Furthermore, the DGLAP equations are formulated within specific flavor schemes, such as the fixed-flavor-number scheme (FFNS) or the variable-flavor-number scheme (VFNS) \cite{Guzzi:2024can,Risse:2025smp,Goharipour:2018dsa,Barontini:2024xgu}. Solving the evolution in the VFNS requires using different equations below and above heavy quark thresholds, with the solutions matched at the thresholds~\cite{deFlorian:2025yar}. Programs such as \texttt{APFEL}~\cite{Bertone:2013vaa} and \texttt{QCDNUM}~\cite{Botje:2010ay} support both FFNS and VFNS evolution.
It is worth noting in this context that the evolution equations involve quarks, antiquarks, and gluons. Due to symmetries in the splitting functions, the equations can be simplified into separate evolution equations for the non singlet quark distributions, along with a coupled system that governs the singlet quark and gluon distributions~\cite{Salam:2008qg,Bertone:2013vaa,Botje:2010ay}.

Now, an important issue to consider is the backward evolution. Typically, the evolution of PDFs is performed forward, starting from a lower input scale $\mu_0^2$ and progressing to a higher scale $\mu^2$. However, backward evolution involves evolving PDFs from a higher scale back to a lower scale. Mathematically, it is the same evolution equation,
but the integration direction for the variable $\ln \mu^2$ is reversed.  This process can be unstable, especially at low energy scales where the strong coupling constant $\alpha_s(\mu^2)$ becomes large. At these scales, the methods of perturbative QCD no longer work well. For this reason, programs like \texttt{QCDNUM} have safety checks and will stop the calculation if $\alpha_s(\mu^2)$ goes beyond a certain limit~\cite{Botje:2010ay}.
Also, another point is that in the FFNS, the number of active flavors $n_f$ is constant, simplifying backward evolution. In the VFNS, $n_f$ changes at flavor thresholds (e.g., charm, bottom), requiring careful handling of discontinuities in $\alpha_s$ and splitting functions during backward evolution.
In the next section, we demonstrate how the breakdown of the quark number sum rules at scales below the initial scale $\mu_0 $ of a given PDF set
affects the results of the QCD analysis of GPDs as an example.

\section{A Challenge in GPDs analysis}\label{sec:three}

GPDs are essential ingredients in the theoretical description of both elastic scattering and hard exclusive processes~\cite{Lorce:2025aqp,Ji:1998pc,Goeke:2001tz,Diehl:2003ny,Ji:2004gf,Belitsky:2005qn,Goloskokov:2007nt,Boffi:2007yc,Guidal:2013rya,Diehl:2015uka,Kumericki:2016ehc,Moutarde:2018kwr,Mezrag:2022pqk,Boer:2025ixc,Guo:2025muf}. At vanishing skewness ($ \xi = 0 $), which corresponds to zero longitudinal momentum transfer, GPDs are directly related to various hadron form factors (FFs), including electromagnetic, axial, gravitational, and transition FFs~\cite{Diehl:2004cx,Diehl:2013xca,Hashamipour:2022noy,Bernard:2001rs,Guidal:2004nd,Polyakov:2018zvc,Goharipour:2024atx,Goharipour:2025lep,Deng:2026gik,Arbabifar:2026tev}. It has also been suggested that GPDs may provide insights relevant to cosmological studies~\cite{Goharipour:2025vtq}.
Recently, the MMGPDs Collaboration has extracted GPDs at $ \xi = 0 $ through a global analysis of experimental data, including electromagnetic FFs and elastic electron-proton scattering cross sections (both differential and reduced)~\cite{Goharipour:2024atx,Goharipour:2024mbk}. Based on their framework, the valence quark GPDs $H_v^q(x, \mu^2, t) $, where $ q $ denotes the quark flavor and $ t $ is the momentum transfer squared, can be parametrized using the following ansatz:
\begin{equation}
H_v^q(x,\mu^2,t) = q_v(x,\mu^2)\exp\left[t f_v^q(x)\right]\,,
\label{Eq4}
\end{equation}
where $ q_v(x,\mu^2) $ represents the unpolarized valence PDFs for flavors $ q = u, d $, with the strange quark contribution neglected. The profile function $ f_v^q(x) $ is determined by fitting to data using a suitable functional form. It is important to note that in the forward limit, where both $ \xi \rightarrow 0 $ and $ t \rightarrow 0 $, the GPDs reduce exactly to the corresponding PDFs. In their analysis, the MMGPDs Collaboration employed the next-to-leading order (NLO) \texttt{NNPDF40} valence quark distributions~\cite{NNPDF:2021njg} evaluated at a scale of $ \mu = 2 \, \mathrm{GeV} $ as the forward limit input. However, the sensitivity of the results to the choice of the scale $ \mu $ has not yet been investigated.

\begin{figure}[!tb]
    \centering
    \includegraphics[scale=0.7]{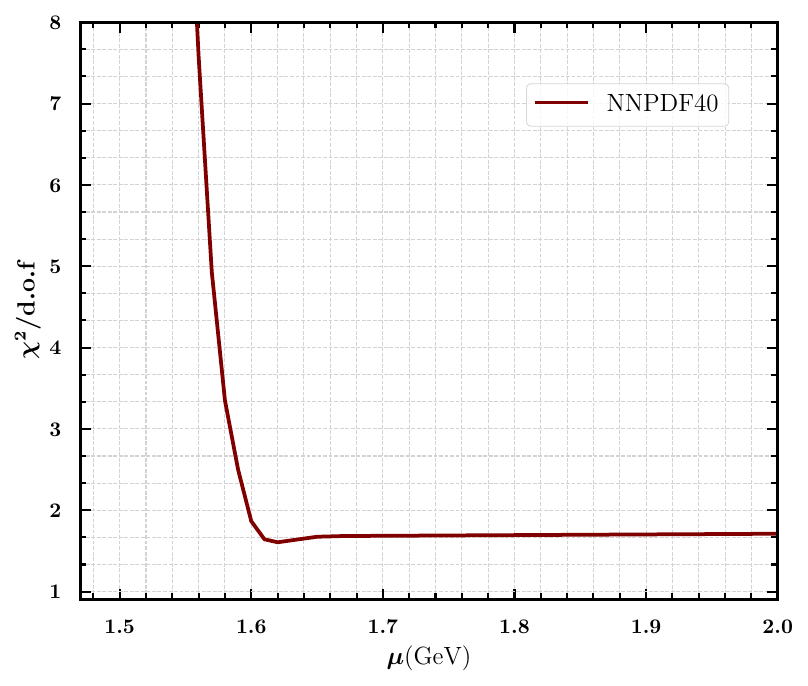}   
    \caption{The total $ \chi^2 $ divided by the number of degrees of freedom, $ \chi^2/\mathrm{d.o.f.} $, as a function of $ \mu $ in which the \texttt{NNPDF40} PDFs set is chosen in the ansatz~(\ref{Eq4}).}
    \label{fig:chi2}
\end{figure}

It is now of interest to repeat the recent analysis by the MMGPDs Collaboration~\cite{Goharipour:2024mbk} while varying the factorization scale $ \mu $, in order to investigate how different choices of $ \mu $ may affect the final results. For further information on the phenomenological framework, data selection, etc., we refer readers to the original papers~\cite{Goharipour:2024atx,Goharipour:2024mbk}.
Figure~\ref{fig:chi2} displays the values of the total $ \chi^2 $ divided by the number of degrees of freedom, $ \chi^2/\mathrm{d.o.f.} $, as a function of $ \mu $ in which the \texttt{NNPDF40} PDF set is used in the ansatz~(\ref{Eq4}). As can be seen, for $ \mu  \lesssim 1.6 \, \mathrm{GeV} $, the value of $ \chi^2/\mathrm{d.o.f.} $ increases sharply, while for larger values of $ \mu $, the curve becomes flat. The flat part suggests that the analysis is relatively insensitive to the scale choice, though taking smaller values of $ \mu $ leads to a slightly smaller total $\chi^2 $. However, the steep increase at lower $ \mu $ indicates a strong sensitivity to the scale choice in this region, suggesting potential limitations in the reliability of the PDFs at such scales.

Upon further investigation, we found that a similar behavior is also observed when using the NLO \texttt{CT18}~\cite{Hou:2019efy} PDFs, although it appears at a different scale, namely around $ \mu \simeq 1.3 \, \mathrm{GeV} $. In contrast, this issue does not arise when employing the NLO \texttt{MSHT20}~\cite{Bailey:2020ooq} PDFs, even down to $ \mu = 1 \, \mathrm{GeV} $. A more detailed investigation revealed that this discrepancy stems from choosing $ \mu $ below the initial scale $ \mu_0 $ of the selected PDF set. Note that
the initial scales $ \mu_0 $ at which the \texttt{NNPDF40} and \texttt{CT18} PDF sets have been parametrized are $ \mu_0 = 1.65 \, \mathrm{GeV} $ and $ \mu_0 = 1.3 \, \mathrm{GeV} $, respectively.

Since the valence GPDs are related to the valence PDFs through Eq.~(\ref{Eq4}), it would be informative to examine the quark number sum rules, which connect the integrated PDFs to the valence quark content of the hadron, as a function of $ \mu $ for the aforementioned PDF sets. For the proton, the valence up and down quark distributions must satisfy the following relations:
\begin{equation}
\int_{0}^{1} dx \, [u(x,\mu^2)- \overline{u}(x,\mu^2)] = 2\,, \quad \int_{0}^{1} dx \, [d(x,\mu^2)- \overline{d}(x,\mu^2)] = 1\,,
\label{Eq5}
\end{equation}
which reflect the proton's quark content of two up quarks and one down quark. These sum rules are preserved under QCD evolution and thus serve as valuable consistency checks. Any violation of these relations could indicate issues such as those observed in the behavior shown in Fig.~\ref{fig:chi2}.

\begin{figure}[!tb]
    \centering
    \includegraphics[scale=0.7]{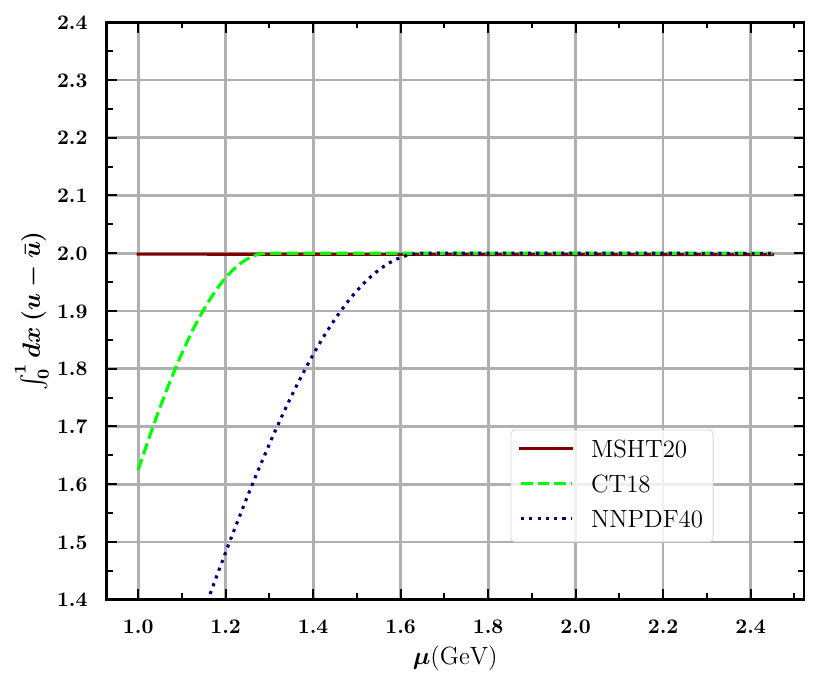}   
    \caption{The quark number sum rule for the up valence PDF of the proton as a function of factorization $ \mu $ for three PDF sets, namely \texttt{NNPDF40}~\cite{NNPDF:2021njg}, \texttt{CT18}~\cite{Hou:2019efy}, and \texttt{MSHT20}~\cite{Bailey:2020ooq}.}
    \label{fig:uv}
\end{figure}
\begin{figure}[!tb]
    \centering
    \includegraphics[scale=0.7]{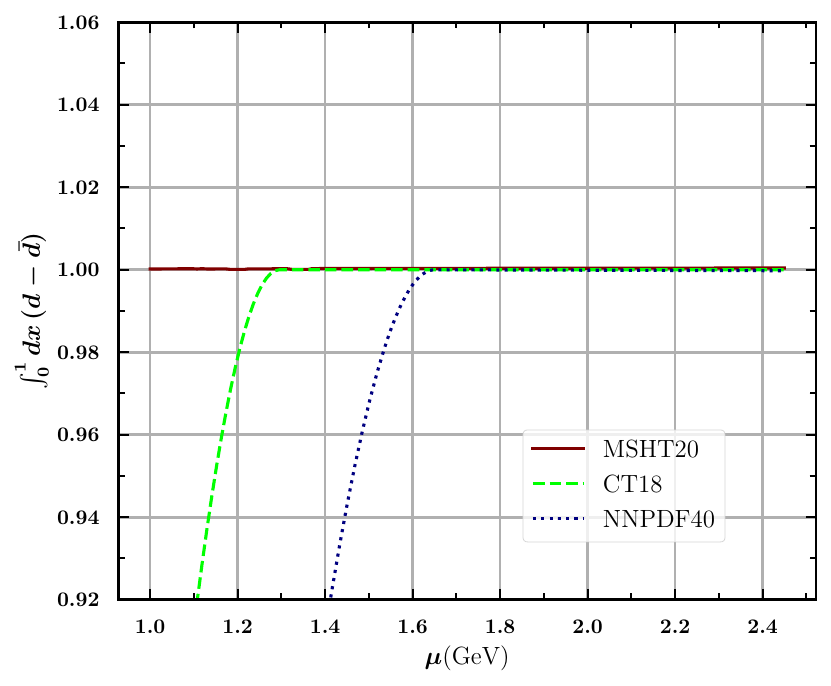}   
    \caption{Same as Fig.~\ref{fig:uv} but for the down valence PDF.}
    \label{fig:dv}
\end{figure}
As illustrated in Figs.~\ref{fig:uv} and~\ref{fig:dv}, our numerical analysis shows that while the quark number sum rules are accurately satisfied at scales above the initial scale $ \mu_0 $ for the \texttt{NNPDF40} and \texttt{CT18} PDF sets, significant deviations from the exact sum rule begin to appear when these PDFs are evaluated at scales below the input parametrization scale $ \mu_0 $. In contrast, no violation is observed for the \texttt{MSHT20} set, which is consistent with the fact that it adopts $ \mu_0 = 1\,\mathrm{GeV} $. This apparent violation does not reflect a breakdown of QCD evolution itself, but rather a limitation of the numerical implementation and interpolation of PDF grids when evaluated below their input parametrization scale. Since the normalization of valence GPDs in Eq.~(\ref{Eq4}) is directly inherited from the valence PDFs, any deviation from the quark number sum rules will propagate into the extracted GPDs. Note that the limitations pertain to backward evolution and the use of PDFs below their input scale are well known within the context of collinear PDF analyses. The aim of the present section was to demonstrate explicitly how these issues manifest themselves within the framework of GPD extractions, where PDFs enter as essential inputs through the forward limit. Our analysis shows that such effects directly propagate into the
extracted GPDs and can therefore impact phenomenological applications.


\section{Determination of GPDs at zero skewness}\label{sec:four}

Motivated by the results obtained in the previous section, we present six QCD analyses of GPDs at zero skewness using three different PDF sets, namely \texttt{NNPDF40}~\cite{NNPDF:2021njg}, \texttt{CT18}~\cite{Hou:2019efy}, and \texttt{MSHT20}~\cite{Bailey:2020ooq}, at both NLO and NNLO. Each PDF set is taken at its corresponding input parametrization scale $\mu_0$ except for \texttt{NNPDF40} which is taken at $ \mu=2 $ GeV. To this end, we adopt the MMGPDs Collaboration framework~\cite{Goharipour:2024atx,Goharipour:2024mbk}, with some minor modifications to both the experimental data selection and the phenomenological framework.

In particular, we exclude the Mainz data~\cite{A1:2013fsc}, which exhibit significant inconsistency with the bulk of other measurements. So, we use the remaining world data, including the $ R^p=\mu_p G_E^p/G_M^p $ polarization, $ G_E^n $, and $ G_M^n/\mu_n G_D $ measurements from the YAHL18 analysis~\cite{Ye:2017gyb}. We also include measurements of $G_E^p/G_D$ from AMT07~\cite{Arrington:2007ux}, the reduced cross section $\sigma_R$ from the JLab experiment~\cite{Christy:2021snt} (GMp12), and differential cross-section measurements from the PRad experiment~\cite{Xiong:2019umf}. Additional details on the formalism and data treatment can be found in Refs.~\cite{Goharipour:2024atx,Goharipour:2024mbk}. Here, $G_E$ and $G_M$ denote the electric and magnetic form factors, respectively, while the superscripts $p$ and $n$ refer to the proton and neutron. The magnetic moment of the nucleon, $ j=p,n $ has been represented by $ \mu_j $ and  $ G_D=(1+Q^2/\Lambda^2)^{-2} $ denotes the dipole form factor with $ \Lambda^2=0.71 $ GeV$ ^2 $.

Let's briefly review the phenomenological framework used in the present study, highlighting the modifications with respect to Refs.~\cite{Goharipour:2024atx,Goharipour:2024mbk}. We adopt the same ansatz for the unpolarized valence GPDs $H_v^q$ and $E_v^q$ as given in Eq.~(\ref{Eq4}). The functional form of the corresponding profile functions, $ f_v^q(x) $ for GPD $ H $ and $ g_v^q(x) $ for GPD $ E $, is parametrized as
\begin{equation}
\label{Eq6}
{\cal F}_v^q(x)=\alpha_q^{\prime}(1-x)^3\log\frac{1}{x}+B_q(1-x)^3 + A_qx(1-x)^2\,.
\end{equation}
The forward limit of the GPD $E_v^q$, denoted by $e_v^q(x)$, is taken in the form
\begin{equation}
\label{Eq7}
e_v^q(x)=\kappa_q N_q x^{-\alpha_q} (1-x)^{\beta_q} \left(1+\gamma_q\sqrt{x}\right)\,,
\end{equation}
following the parametrization used in the MMGPDs Collaboration analyses~\cite{Goharipour:2024atx,Goharipour:2024mbk}. Here, the anomalous magnetic moments of the up and down quarks are fixed to $\kappa_u=1.67$ and $\kappa_d=-2.03$, respectively, in the units of nuclear magneton. The normalization constant $N_q$ is determined by the condition
\begin{equation}
\label{Eq8}
\int_0^1 dx\, e_v^q(x)=\kappa_q\,.
\end{equation}

The electric and magnetic form factors, $ G_E(t)= F_1(t) + \frac{t}{4m^2} F_2(t)$, and $ G_M(t)= F_1(t)+ F_2(t) $, where $ F_1 $ and $ F(2) $ are the Dirac and Pauli FFs, respectively, and $ m $ represents the nucleon mass, can be calculated, e.g, for the proton, using the following relations
\begin{align}
F^p_1(t)=\sum_q e_q \int_{0}^1 dx\, H_v^q(x,\mu^2,t)\,, \nonumber \\ 
F^p_2(t)=\sum_q e_q \int_{0}^1 dx\, E_v^q(x,\mu^2,t)\,,
\label{Eq9}
\end{align}
where $ e_q $ refers to the electric charge of the constituent quark $ q $.

We make several modifications in the treatment of the fit parameters. In particular, we fix $\beta_u=4.65$, $\beta_d=5.25$, $\gamma_u=4$, $\gamma_d=0$, and impose the constraint $\alpha_d=\alpha_u$, following the \texttt{DK13} analysis~\cite{Diehl:2013xca}. For the profile functions, we assume $\alpha^{\prime}_{f_v^u} - \alpha^{\prime}_{f_v^d}=0.2$, which differs slightly from the value of $0.1$ used in Ref.~\cite{Diehl:2013xca}. Furthermore, from a parametrization scan, we find that the optimal fit is obtained by imposing the relations $\alpha^{\prime}_{g_v^u}=\alpha^{\prime}_{f_v^u}$ and $\alpha^{\prime}_{g_v^d}=\alpha^{\prime}_{f_v^d}$. 
These constraints reduce the number of free parameters and improve the stability of the fit.
Note that releasing these parameters does not lead to a significant change in the values of $ \chi^2/\mathrm{d.o.f.} $, but induces strong correlations between the parameters of the GPDs $ H $ and $ E $, as well as among the parameters of the GPD $ E $ themselves.

As in Refs.~\cite{Goharipour:2024atx,Goharipour:2024mbk}, the minimization procedure is performed using the \texttt{MINUIT} package from the CERN program library~\cite{James:1975dr}, which allows for a reliable determination of the optimal values of the free parameters. The uncertainties on the extracted parameters are estimated using the standard Hessian method~\cite{Pumplin:2001ct} with the tolerance criterion $\Delta \chi^2 = 1$. In addition, we impose the condition $g_v^q(x) < f_v^q(x)$, which is implemented directly within the fit procedure, in order to ensure the positivity property of GPDs over a broad range of $x$. Further details on the phenomenological framework and fitting strategy can be found in Refs.~\cite{Goharipour:2024atx,Goharipour:2024mbk}.

\begin{table}[!tb]
\scriptsize
\setlength{\tabcolsep}{5pt}
\renewcommand{\arraystretch}{1.4}
\caption{The values of $ \chi^2 $ per number of data points obtained from the six present analyses using the \texttt{NNPDF40}~\cite{NNPDF:2021njg}, \texttt{CT18}~\cite{Hou:2019efy}, and \texttt{MSHT20}~\cite{Bailey:2020ooq} PDF sets at NLO and NNLO accuracy. The values of total $ \chi^2 $ divided by the number of degrees of freedom are presented in the last row. Note that we have taken the \texttt{NNPDF40}, \texttt{CT18}, and \texttt{MSHT20} PDF sets at $ \mu=2, 1.3 $ and $ 1 $ GeV, respectively.}
\label{tab:chi2}
\begin{tabular}{lcccccc}
\hline
\hline
Observable & \multicolumn{6}{c}{$\chi^2 / N_{\textrm{pts.}}$} \\
\cline{2-7}
 & NNPDF40 NLO & NNPDF40 NNLO & CT18 NLO & CT18 NNLO & MSHT20 NLO & MSHT20 NNLO \\
\hline
\hline
AMT07 $G_E^p/G_D$~\cite{Arrington:2007ux} & $51.4 / 47$ & $66.1 / 47$ & $68.0 / 47$ & $67.5 / 47$ & $63.5 / 47$ & $68.4 / 47$ \\
YAHL18 $R^p$~\cite{Ye:2017gyb}            & $114.8 / 69$ & $119.0 / 69$ & $111.8 / 69$ & $110.6 / 69$ & $111.4 / 69$ & $113.5 / 69$ \\
YAHL18 $G_E^n$~\cite{Ye:2017gyb}          & $25.8 / 38$ & $25.7 / 38$ & $25.3 / 38$ & $25.9 / 38$ & $25.8 / 38$ & $26.2 / 38$ \\
YAHL18 $G_M^n/\mu_n G_D$~\cite{Ye:2017gyb}& $45.9 / 33$ & $46.4 / 33$ & $44.2 / 33$ & $43.9 / 33$ & $44.6 / 33$ & $45.7 / 33$ \\
GMp12 $\sigma_R$~\cite{Christy:2021snt}   & $15.5 / 13$ & $19.7 / 13$ & $15.8 / 13$ & $15.3 / 13$ & $14.9 / 13$ & $15.3 / 13$ \\
PRad $\dv{\sigma}{\Omega}$ 1.1 GeV~\cite{Xiong:2019umf} & $9.7 / 33$ & $9.2 / 33$ & $8.9 / 33$ & $8.9 / 33$ & $8.8 / 33$ & $8.5 / 33$ \\
PRad $\dv{\sigma}{\Omega}$ 2.2 GeV~\cite{Xiong:2019umf} & $42.5 / 38$ & $42.3 / 38$ & $43.4 / 38$ & $43.5 / 38$ & $66.2 / 38$ & $61.1 / 38$ \\
\hline
Total $\chi^2 /\mathrm{d.o.f.}$ & $305.6 / 261$ & $328.4 / 261$ & $317.4 / 261$ & $315.6 / 261$ & $335.2 / 261$ & $338.7 / 261$ \\
\hline
\hline
\end{tabular}
\end{table}
In Table~\ref{tab:chi2}, we have summarized the quality of the fits obtained in the present analyses in terms of the partial and total values of $\chi^2$. The table lists the contributions to $\chi^2$ from each individual experimental dataset, divided to the corresponding number of data points, as well as the total $\chi^2$ divided by the number of degrees of freedom (which is the same for all analyses, i.e., $ \mathrm{d.o.f.}=261 $).

A comparison of the different datasets allows one to assess the consistency of the phenomenological framework and the impact of the selected experimental inputs. In particular, the exclusion of the Mainz $G_E^p$ data leads to a noticeable improvement in the overall fit quality, as reflected in the reduced total $\chi^2/\mathrm{d.o.f.}$ (it reduced from $ 569.9 / 333 $ in Ref.~\cite{Goharipour:2025yxm} to $305.6 / 261$ in the present study). The remaining datasets, including the world polarization measurements, neutron form factor data, and cross-section measurements from the GMp12 and PRad experiments, are well described within the present framework.

Overall, the values of $\chi^2/N_{\mathrm{pts.}}$ indicate a good agreement between the theoretical parametrization and the experimental data which confirms the stability and reliability of our extracted GPDs at zero skewness. The results show that using the \texttt{NNPDF40} PDFs~\cite{NNPDF:2021njg} at NLO with $\mu = 2~\mathrm{GeV}$ leads to the smallest total $\chi^2$. When using the NNLO version of \texttt{NNPDF40}, the total $\chi^2$ increases by about 23 units due to a moderately worse description of the  AMT07~\cite{Arrington:2007ux} data within the present framework.

In contrast, the fits obtained using the \texttt{CT18} and \texttt{MSHT20} PDF sets show comparable $\chi^2$ values at both NLO and NNLO accuracy. This suggests a relatively weak dependence on the perturbative order for these PDF inputs. The largest $\chi^2$ values are obtained when using the \texttt{MSHT20} PDFs with $\mu = 1~\mathrm{GeV}$. As seen in Table~\ref{tab:chi2}, this behavior is mainly driven by the PRad data~\cite{Xiong:2019umf}, which cover the low-$t$ region ($0.00022 \le -t \le 0.05819~\mathrm{GeV}^2$) and appear to be less well described by the \texttt{MSHT20}-based fits. A more detailed comparison between the different PDF sets and perturbative orders will be presented in the following.

\begin{widetext}
\begin{center}
\begin{table}[!tb]
\scriptsize
\setlength{\tabcolsep}{6pt}
\renewcommand{\arraystretch}{1.4}
\caption{The optimum values of the parameters of the profile functions $ f_v^q(x) $ and $ g_v^q(x) $ in Eq.~(\ref{Eq6}) and the forward limit distributions $e_v^q(x)$ in Eq.~(\ref{Eq7}), obtained from the six analyses performed in this work using the \texttt{NNPDF40}~\cite{NNPDF:2021njg}, \texttt{CT18}~\cite{Hou:2019efy}, and \texttt{MSHT20}~\cite{Bailey:2020ooq} PDF sets which have been taken at $ \mu=2, 1.3 $ and $ 1 $ GeV, respectively. Note that we have considered $\beta_u=4.65$, $\beta_d=5.25$, $\gamma_u=4$, $\gamma_d=0$,  $\alpha_d=\alpha_u$, $\alpha^{\prime}_{f_v^u} - \alpha^{\prime}_{f_v^d}=0.2$, $\alpha^{\prime}_{g_v^u}=\alpha^{\prime}_{f_v^u}$ and $\alpha^{\prime}_{g_v^d}=\alpha^{\prime}_{f_v^d}$. }
\label{tab:par}
\begin{tabular}{lccccccc}
\hline
\hline
Distribution & Parameter 
& NNPDF40 NLO 
& NNPDF40 NNLO 
& CT18 NLO 
& CT18 NNLO 
& MSHT20 NLO 
& MSHT20 NNLO \\
\hline
\hline

$f_v^u(x)$
& $\alpha^{\prime}$ 
&  $ 0.717 \pm 0.019 $ & $ 0.834 \pm 0.022 $ & $ 0.988 \pm 0.024 $ & $ 1.007 \pm 0.024 $ & $ 0.945 \pm 0.025 $ & $ 1.036 \pm 0.028 $ \\
& $A$ 
&  $ 0.963 \pm 0.038 $ & $ 1.089 \pm 0.043 $ & $ 1.492 \pm 0.040 $ & $ 1.458 \pm 0.039 $ & $ 1.691 \pm 0.040 $ & $ 1.727 \pm 0.043 $ \\
& $B$ 
&  $ 0.863 \pm 0.047 $ & $ 0.665 \pm 0.053 $ & $ 0.567 \pm 0.054 $ & $ 0.603 \pm 0.053 $ & $ 0.748 \pm 0.056 $ & $ 0.712 \pm 0.062 $ \\

\hline
$f_v^d(x)$
& $A$ 
&  $ 2.870 \pm 0.277 $ & $ 2.904 \pm 0.269 $ & $ 2.718 \pm 0.249 $ & $ 2.661 \pm 0.243 $ & $ 3.223 \pm 0.246 $ & $ 3.561 \pm 0.252 $ \\
& $B$ 
&  $ 1.005 \pm 0.062 $ & $ 0.815 \pm 0.065 $ & $ 0.941 \pm 0.067 $ & $ 0.972 \pm 0.065 $ & $ 0.993 \pm 0.070 $ & $ 0.773 \pm 0.076 $ \\

\hline
$g_v^u(x)$
& $A$ 
&  $ 0.897 \pm 0.070 $ & $ 0.879 \pm 0.066 $ & $ 1.106 \pm 0.077 $ & $ 1.152 \pm 0.080 $ & $ 1.075 \pm 0.069 $ & $ 1.102 \pm 0.067 $ \\
& $B$ 
&  $ 0.415 \pm 0.046 $ & $ 0.454 \pm 0.056 $ & $ 0.464 \pm 0.068 $ & $ 0.464 \pm 0.069 $ & $ 0.416 \pm 0.062 $ & $ 0.440 \pm 0.072 $ \\

\hline
$g_v^d(x)$
& $A$ 
&  $ 2.111 \pm 0.266 $ & $ 2.616 \pm 0.275 $ & $ 2.491 \pm 0.264 $ & $ 2.430 \pm 0.258 $ & $ 2.551 \pm 0.261 $ & $ 2.791 \pm 0.268 $ \\
& $B$ 
&  $-0.047 \pm 0.072 $ & $-0.257 \pm 0.083 $ & $-0.352 \pm 0.095 $ & $-0.348 \pm 0.096 $ & $-0.352 \pm 0.090 $ & $-0.463 \pm 0.100 $ \\

\hline
$e_v^u(x)$
& $\alpha$ 
&  $ 0.727 \pm 0.012 $ & $ 0.668 \pm 0.015 $ & $ 0.576 \pm 0.020 $ & $ 0.562 \pm 0.020 $ & $ 0.610 \pm 0.018 $ & $ 0.557 \pm 0.022 $ \\
\hline
\hline
\end{tabular}
\end{table}
\end{center}
\end{widetext}
The optimum values of the fit parameters obtained from the six analyses have been summarized in Table~\ref{tab:par}. Overall, the extracted parameters are well constrained, with relatively small uncertainties which indicates a stable and reliable determination of the GPDs within the present framework. We observe that the parameters of the profile functions $f_v^u(x)$ and $f_v^d(x)$ show a moderate dependence on the choice of the input PDF set and perturbative order. In particular, the parameter $\alpha^{\prime}$ of $f_v^u(x)$ tends to increase when going from NLO to NNLO which reflects a slightly stronger $t$ dependence of the GPDs at higher perturbative accuracy.

The parameters related to the $g_v^q(x)$ profile functions are generally less sensitive to the perturbative order. However we can see some dependence on the PDF set, especially for the down-quark sector. The forward limit parameter $\alpha$ of the distributions $e_v^q(x)$ (note that $\alpha_d=\alpha_u$) shows a clear dependence on the input PDF set, with larger values obtained for \texttt{NNPDF40} and smaller values for \texttt{CT18} and \texttt{MSHT20}. This behavior reflects the sensitivity of the extracted GPDs to the shape of the valence PDFs at the corresponding initial scales.

Despite these variations, the overall consistency of the extracted parameters between all six analyses indicates the robustness of the present phenomenological framework. The observed differences remain within a physically reasonable range and do not lead to significant changes in the quality of the fits. This confirms that the extracted GPDs are stable with respect to both the choice of PDF set and perturbative order.

\begin{figure}[!tb]
    \centering
\includegraphics[scale=0.5]{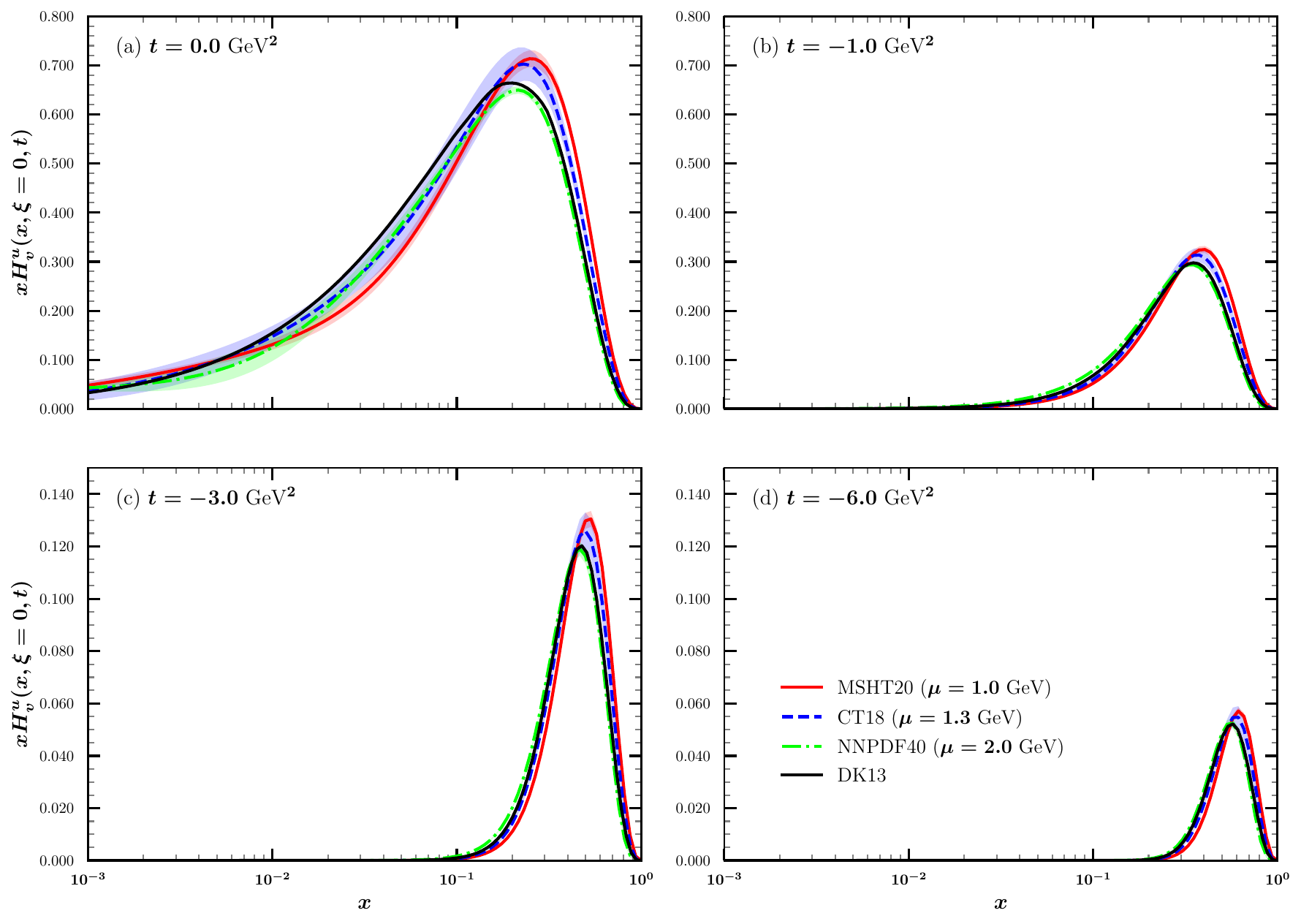}   
    \caption{A comparison between our results for the GPD $ xH_v^u(x) $ and those from the \texttt{DK13} analysis~\cite{Diehl:2013xca} at four momentum transfer values: (a) $ t = 0 $, (b) $ t = -1 $, (c) $ t = -3 $, and (d) $ t = -6~\mathrm{GeV}^2 $. See the text for more details.}
\label{fig:Huv}
\end{figure}
A comparison between our NLO results for the GPD $ xH_v^u(x) $ and the \texttt{DK13} analysis~\cite{Diehl:2013xca} at four momentum transfer values, $ t = 0, -1, -3, $ and $ -6~\mathrm{GeV}^2 $, is presented in Fig.~\ref{fig:Huv}. Note that at $ t=0 $ each GPD set reduces to its corresponding PDF set. The corresponding PDF uncertainties have also been included in the error bands in this figure as well as the other plots which are presented in the following. It should be noted that our results correspond to different scales $ \mu $; $ \mu=2, 1.3 $ and $ 1 $ GeV for the \texttt{NNPDF40}~\cite{NNPDF:2021njg}, \texttt{CT18}~\cite{Hou:2019efy}, and \texttt{MSHT20}~\cite{Bailey:2020ooq} PDF sets, respectively.  In addition to the expected suppression of GPDs with increasing $ |t| $, the figure clearly demonstrates the effects of scale evolution from $ \mu = 1 $ to $ \mu = 2~\mathrm{GeV} $. As expected, increasing $ \mu $ reduces the overall magnitude of the distribution and shifts its peak toward smaller values of $ x $. Note that at larger values of $ |t| $, different GPD sets became more consistent. We do not present a comparison at NNLO, as the corresponding results exhibit similar behavior. Among our NLO results, the one based on the \texttt{NNPDF40} PDF set is directly comparable to the \texttt{DK13} result, since both correspond to $ \mu = 2~\mathrm{GeV} $. As shown, these results display good consistency, even though \texttt{DK13} employed the \texttt{ABM}~\cite{Alekhin:2012ig} PDF set.

\begin{figure}[!htb]
    \centering
\includegraphics[scale=0.5]{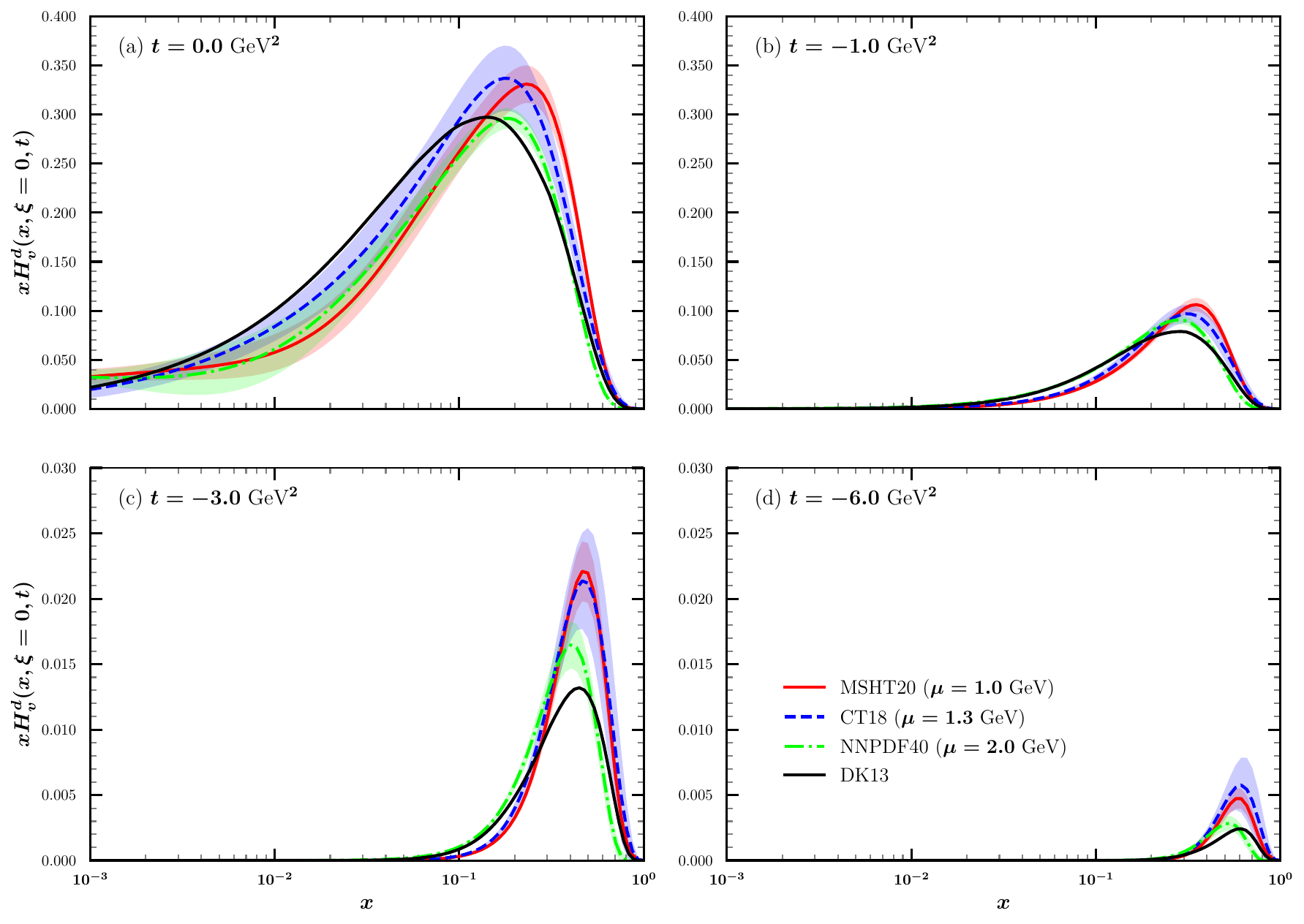}    
    \caption{Same as Fig.~\ref{fig:Huv} but for GPD $ xH_v^d(x) $. }
\label{fig:Hdv}
\end{figure}
Figure~\ref{fig:Hdv} displays the same results as Fig.~\ref{fig:Huv}, but for the GPD $ xH_v^d(x) $. Here, the differences among our results, as well as those with respect to the \texttt{DK13} analysis, are more pronounced compared to the case of $ xH_v^u(x) $ shown in Fig.~\ref{fig:Huv}, even at larger values of $ |t| $. Although the discrepancies between the forward limits of the GPDs (PDFs) play an important role in this case, it is also evident from the figure that the $ t $ dependence of our extracted GPDs differs to some extent. In particular, the GPD set based on the \texttt{CT18} PDFs undergoes less suppression with increasing $ |t| $ compared to the other GPD sets. By comparing Figs.~\ref{fig:Huv} and~\ref{fig:Hdv}, one finds that the down-quark GPD is suppressed more strongly than the up-quark GPD as $ |t| $ increases. Upon comparison with the \texttt{DK13} results, one observes that the \texttt{DK13} distribution shifts more toward the large-$ x $ region with increasing $ |t| $. Moreover, the \texttt{DK13} distribution exhibits greater suppression than our GPDs at higher $ |t| $. These findings clearly indicate that the available elastic electron scattering data impose weaker constraints on the GPD $ xH_v^d(x) $ than on $ xH_v^u(x) $.

\begin{figure}[!tb]
    \centering
\includegraphics[scale=0.5]{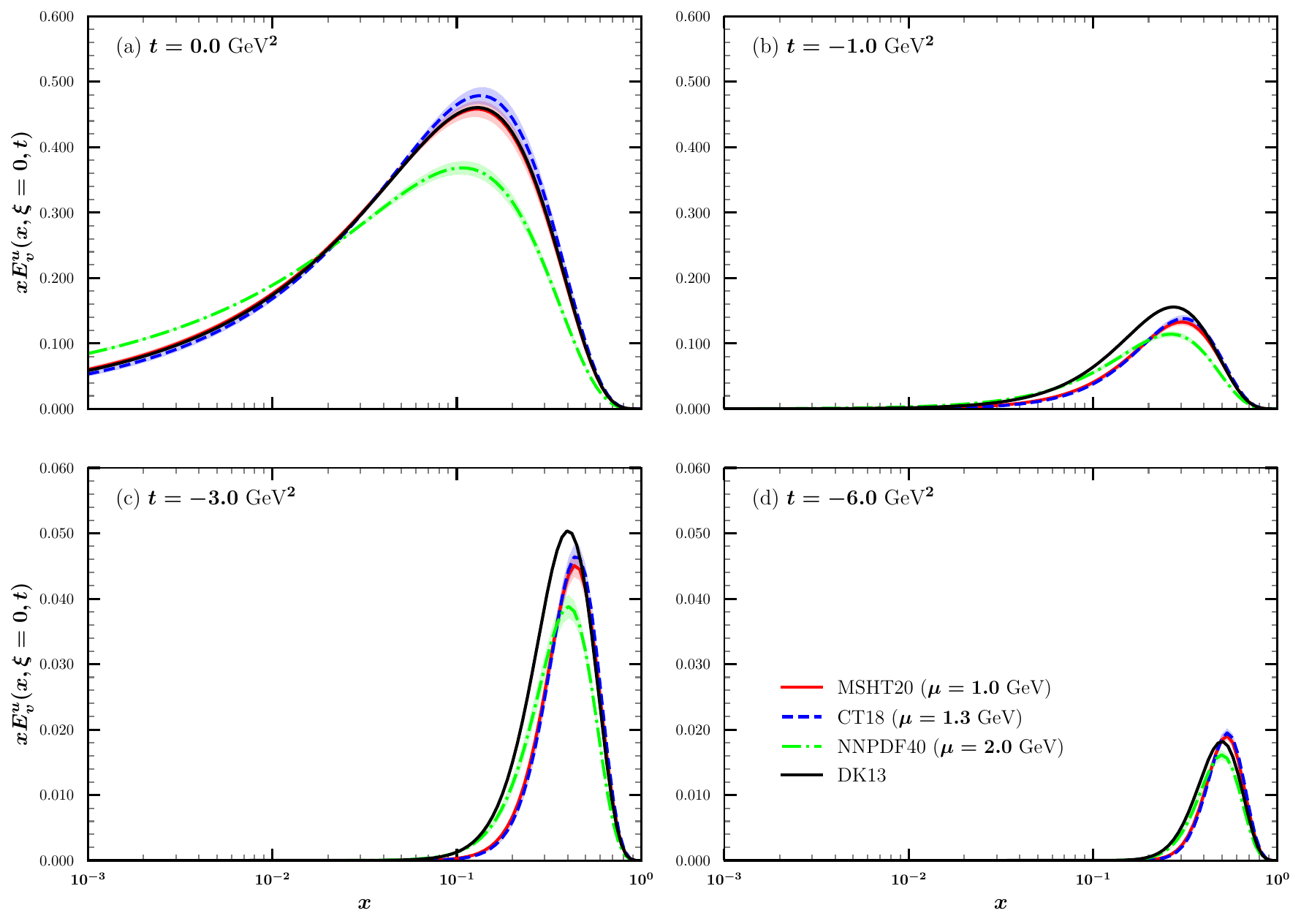}    
    \caption{Same as Fig.~\ref{fig:Huv} but for GPD $ xE_v^u(x) $. }
\label{fig:Euv}
\end{figure}
Our results for GPD $ xE_v^u(x) $ are shown in Fig.~\ref{fig:Euv} and compared again with \texttt{DK13} results at $ t=0,-1,-3,-6 $ GeV$ ^2 $. As can be seen, among the different GPD sets, \texttt{CT18} and \texttt{MSHT20}, corresponding to scales $ \mu=1.3 $ GeV and $ \mu=1 $ GeV respectively, exhibit better consistency across all values of $ |t| $.  
In this case, a remarkable discrepancy exists between our \texttt{NNPDF40}-based GPD and the \texttt{DK13} results, despite both being evaluated at $ \mu=2 $ GeV. This is particularly interesting considering the fact that the \texttt{NNPDF40} GPD set at NLO yields a lower $ \chi^2 $ (see Table~\ref{tab:chi2}) which indicates a better description of the data among all six analyses performed in this work.  
This finding motivates a more detailed investigation into the contribution of the down quark within the nucleon structure and could potentially prompt a revision of its role in nucleon properties, especially at $ t=0 $.  
It is worth noting that at $ t=0 $, all GPD sets are in excellent agreement except for the \texttt{NNPDF40}-based GPD. However, as $ |t| $ increases, the GPDs become more similar though they correspond to different values of $ \mu $.

\begin{figure}[!htb]
    \centering
\includegraphics[scale=0.5]{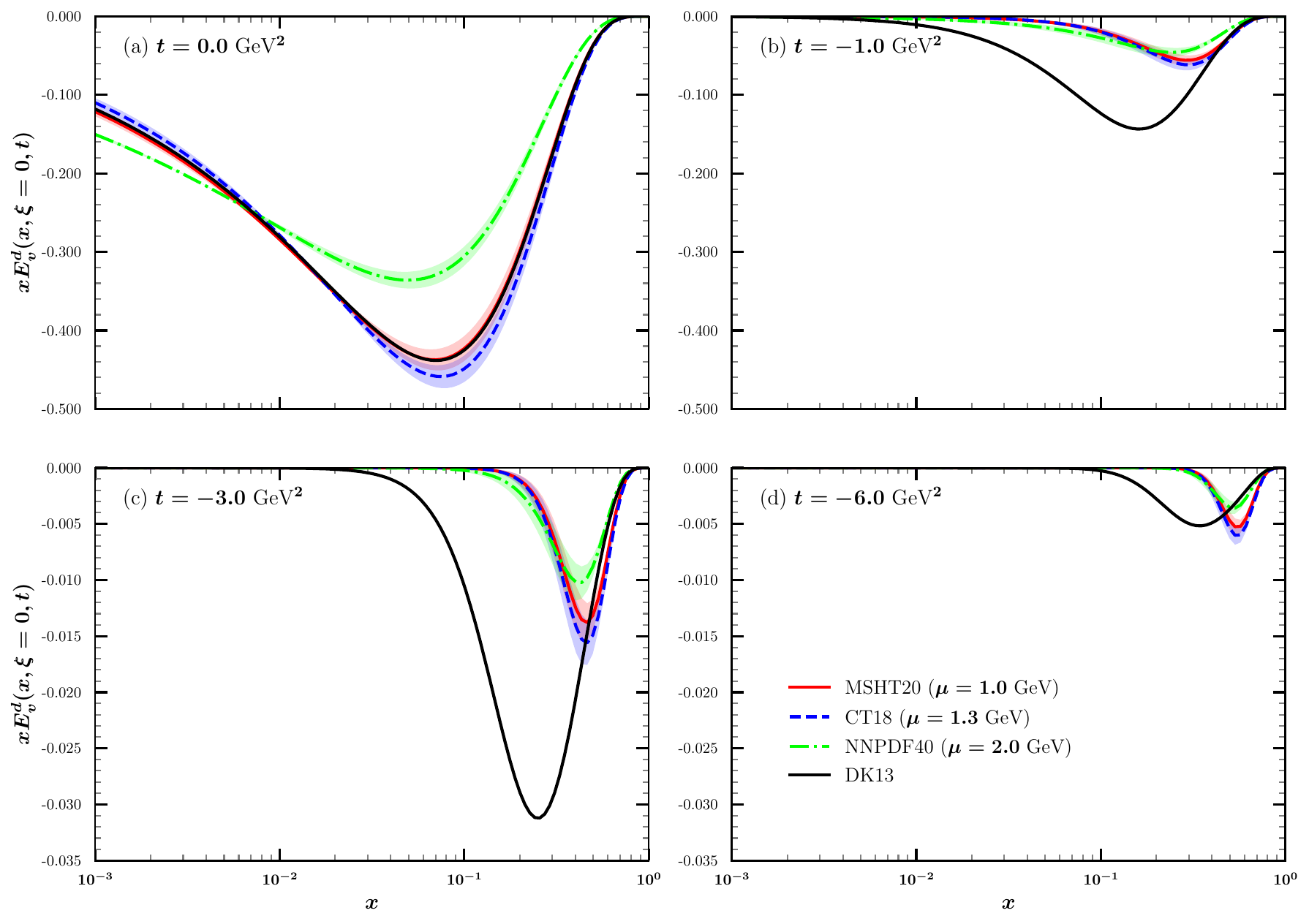}
    \caption{Same as Fig.~\ref{fig:Huv} but for GPD $ xE_v^d(x) $. }
\label{fig:Edv}
\end{figure}
Figure~\ref{fig:Edv} presents the same results as Fig.~\ref{fig:Euv}, but for the GPD $ xE_v^d(x) $. In this case, while the GPDs based on the \texttt{CT18} and \texttt{MSHT20} PDFs again exhibit good consistency, the \texttt{NNPDF40}-based GPD remains distinct even at larger values of $ |t| $. The most significant observation to be drawn from this figure is the significant discrepancy between our results and the \texttt{DK13} analysis as $ |t| $ increases, despite the good agreement at $ t=0 $. Overall, our results suggest a more pronounced suppression with increasing $ |t| $. Furthermore, the \texttt{DK13} distribution tends to the smaller-$ x $ region compared with our extracted GPDs. Similar to the behavior observed for the GPD $ H_v^q $, $ xE_v^d(x) $ undergoes greater suppression than $ xE_v^u(x) $ as $ |t| $ increases. These findings further emphasize the critical need for a deeper investigation into the down-quark distribution within the nucleon.

It should be noted that the differences observed here may significantly impact the outcomes of related phenomenological studies concerning hadron structure when utilizing different GPD sets. This suggests that future high-precision measurements could be sensitive to the choice of PDF input used in GPD extractions. As an illustrative example, we calculate the proton charge and magnetic radii, which have attracted considerable attention in the literature~\cite{Pohl:2010zza,Karr:2020wgh,Gao:2021sml,Xiong:2023zih,Chen:2023dxp,Goharipour:2024mbk,Goharipour:2025yxm,Qattan:2026piw}, especially in connection with the proton radius puzzle. We use the six different GPD sets extracted in the present work in order to investigate how the observed differences among the GPDs propagate into the determination of proton radii. This can be very informative, considering the fact that the mean squared of the charge and magnetic radii are obtained from the slopes of the electromagnetic FFs at $ t \rightarrow 0 $,
\begin{align}
\left<r_{pE}^2\right>= \left.  6 \dv{G_E^p}{t} \right|_{t=0} \,, ~~~~~~
\left<r_{pM}^2\right>= \left.  \frac{6}{\mu_p} \dv{G_M^p}{t} \right|_{t=0}\,,
\label{Eq10}
\end{align}
and hence are sensitive to the product of the forward limits (PDFs) and profile functions according to Eqs.~(\ref{Eq4}) and~(\ref{Eq9}).

\begin{figure}[!htb]
    \centering
\includegraphics[scale=0.9]{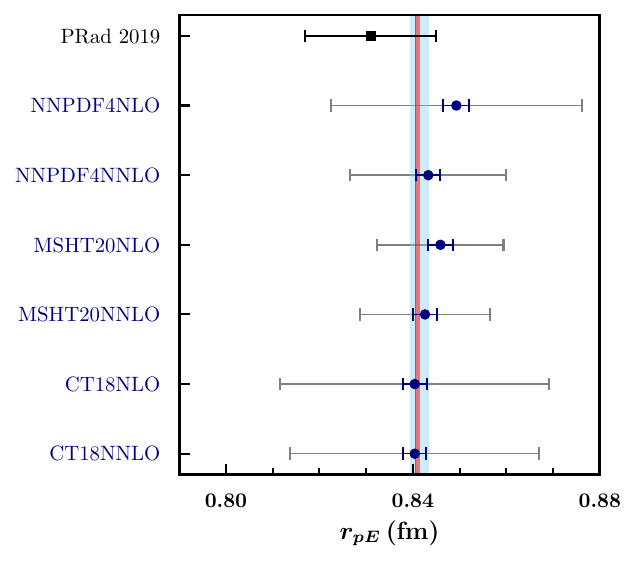}
    \caption{The proton charge radius $r_{pE}$ obtained using the six GPD sets extracted in the present work. The square symbol represents the result from the PRad experiment~\cite{Xiong:2019umf}. The narrow and wide shaded bands correspond to the values reported by the PDG~\cite{ParticleDataGroup:2024cfk} and CODATA~\cite{Tiesinga:2021myr}, respectively. The inner error bars indicate the uncertainties associated with the GPD parametrization, while the outer error bars additionally include the propagated PDF uncertainties.}
\label{fig:rEp}
\end{figure}
Figure~\ref{fig:rEp} presents the results obtained for the proton charge radius $r_{pE}$ in comparison with the corresponding result from the PRad experiment~\cite{Xiong:2019umf} (square symbol), as well as the values reported by the Particle Data Group (PDG), shown as the vertical narrow band~\cite{ParticleDataGroup:2024cfk}, and CODATA, shown as the vertical wider band~\cite{Tiesinga:2021myr}. The inner error bars represent the uncertainties originating from the GPD parametrization only, while the outer error bars additionally include the propagated PDF uncertainties.
As can be seen, all extracted results are mutually consistent within uncertainties. The GPD sets constructed using the \texttt{CT18} PDFs, at both NLO and NNLO accuracy, exhibit better agreement with the PDG and CODATA values, although they are associated with relatively larger uncertainties compared to the other GPD sets. In contrast, the GPD sets based on \texttt{MSHT20} and \texttt{NNPDF40} PDFs tend to yield somewhat larger values of $r_{pE}$, especially in the case of \texttt{NNPDF40} at NLO.

\begin{figure}[!htb]
    \centering
\includegraphics[scale=0.9]{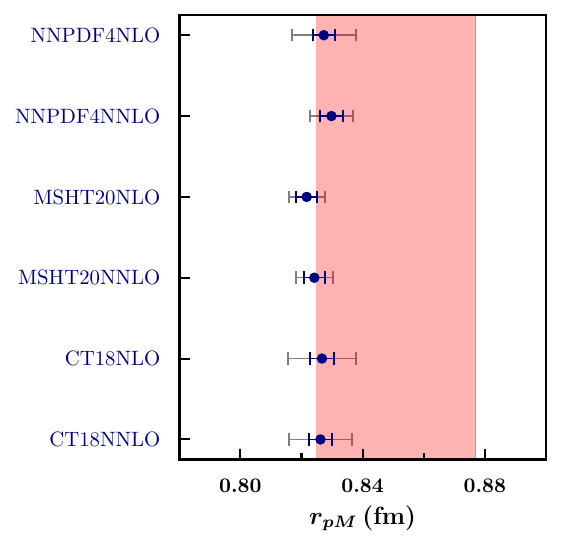}
    \caption{The proton magnetic radius $r_{pM}$ obtained using the six GPD sets extracted in the present work. The vertical band represents the value reported by PDG~\cite{ParticleDataGroup:2024cfk}. The inner error bars correspond to the uncertainties associated with the GPD parametrization, while the outer error bars additionally include the propagated PDF uncertainties.}
\label{fig:rMp}
\end{figure}
In Fig.~\ref{fig:rMp}, we compare the results obtained for the proton magnetic radius, $r_{pM}$, with the corresponding value reported by the PDG, shown as the vertical band. In this case, all extracted results are mutually consistent within uncertainties and are also compatible with the PDG value, although all GPD sets tend to yield smaller central values. Overall, the GPD sets based on the \texttt{NNPDF40} and \texttt{CT18} PDFs exhibit better agreement with the PDG result and with each other, while the \texttt{MSHT20}-based GPD sets, particularly the NLO one, lead to somewhat smaller values of $r_{pM}$.
It is worth noting in this context that the proton magnetic radius, which is related to the magnetic form factor $G_M$, is sensitive to both products of $ f_v(x)\, q_v(x) $ and $ g_v(x)\, e_v(x) $, whereas the proton charge radius $r_{pE}$, related to the electric form factor $G_E$, is sensitive to the product of $ f_v(x)\, q_v(x) $.

According to the results obtained in this section, the differences observed among the six GPD extractions are not only reflected in the fit quality, but may also have implications for physical observables sensitive to GPDs. In particular, variations in the extracted distributions can affect predictions for electromagnetic FFs, nucleon radii, and other related quantities. This suggests that the choice of input PDF set and perturbative order may become relevant in precision studies, where experimental uncertainties are sufficiently small to probe such effects. Therefore, a consistent and physically motivated selection of PDFs is essential for reliable GPD phenomenology.


\section{Summary and conclusions}\label{sec:five}

In the present work, we have investigated the dependence of generalized GPDs at zero skewness on the choice of the input collinear PDFs, perturbative order, and factorization scale. In particular, we studied the impact of employing different modern PDF sets, namely \texttt{NNPDF40}~\cite{NNPDF:2021njg}, \texttt{CT18}~\cite{Hou:2019efy}, and \texttt{MSHT20}~\cite{Bailey:2020ooq}, within the framework of GPD extractions based on elastic electron--nucleon scattering data.

As a first step, we revisited the issue of backward evolution in PDF analyses and investigated how the use of PDFs below their nominal input scale $\mu_0$ manifests itself in the context of GPD phenomenology. Although the limitations associated with backward evolution are well known in collinear PDF analyses, our study demonstrates explicitly that these effects can propagate into GPD extractions through the forward limit. In particular, we showed that employing PDF sets at scales below their initial parametrization scale can lead to a strong deterioration of the fit quality and to violations of the quark number sum rules. These observations highlight the importance of respecting the valid evolution range of PDF sets in phenomenological studies involving GPDs and related observables.

Motivated by these findings, we performed six global analyses of GPDs at zero skewness using the \texttt{NNPDF40}, \texttt{CT18}, and \texttt{MSHT20} PDF sets at both NLO and NNLO accuracy at $\mu = 2$, $1.3$, and $1~\mathrm{GeV}$, respectively. All analyses were carried out using the same experimental data and within a common phenomenological framework, allowing for a direct comparison of the impact of the PDF input and perturbative order on the extracted GPDs.
A comparison of the six analyses shows that the best overall description of the data is achieved using the \texttt{NNPDF40} PDFs at NLO accuracy. The inclusion of NNLO corrections does not lead to a systematic improvement in the fit quality, and in some cases leads to slightly larger values of $\chi^2$. The fits obtained using the \texttt{CT18} PDFs show a comparable level of agreement with the data, while the \texttt{MSHT20} PDFs generally yield somewhat larger $\chi^2$ values, mainly due to their less accurate description of the low-$t$ PRad data. Nevertheless, all six analyses provide an acceptable overall description of the experimental measurements which indicates the stability of the present extraction framework.

Our results indicate that the extracted GPDs exhibit a non-negligible sensitivity to the choice of the input PDF set, particularly in the small-$t$ region. In contrast, the dependence on the perturbative order is found to be relatively mild, suggesting that higher-order QCD corrections do not introduce significant instabilities in the extraction procedure. Regarding the $t$ dependence of the extracted GPDs, two general observations can be made. First, although the six GPD sets correspond to different PDF inputs and perturbative orders, they become increasingly consistent at larger values of $|t|$, particularly for the up-quark GPDs. Second, the down-quark GPDs exhibit a stronger suppression with increasing $|t|$ compared to the up-quark distributions.

In order to investigate the phenomenological implications of the observed PDF dependence, we additionally calculated the proton charge and magnetic radii using the six extracted GPD sets. We found that the differences among the GPD parametrizations propagate into measurable quantities such as nucleon radii, leading to non-negligible variations in the extracted central values and uncertainties. In particular, the GPD sets based on \texttt{CT18} PDFs show better agreement with the PDG and CODATA values for the proton charge radius, while the \texttt{NNPDF40}-based sets yield somewhat larger values. Similarly, the proton magnetic radius exhibits a visible sensitivity to the choice of PDF input. These results demonstrate that the PDF dependence observed in GPD extractions is not merely a technical issue, but can have direct phenomenological consequences for physical observables sensitive to hadron structure. To facilitate the use of our results in phenomenological applications, we provide a public code implementing the extracted GPD sets, enabling the evaluation of $H_v^q(x,t)$ and $E_v^q(x,t)$ at arbitrary values of $x$ and $t$. The package is publicly available at~\cite{MMGPDsData}, together with example script.

The six extracted GPD sets presented in this work provide a useful framework for future theoretical and phenomenological studies of hadron structure at zero skewness, including applications to proton tomography, nucleon form factors, and the investigation of mechanical properties of hadrons. Moreover, the availability of multiple GPD parametrizations based on different modern PDF inputs and perturbative orders offers additional flexibility for future analyses and may help quantify theoretical uncertainties associated with the choice of collinear PDFs in GPD phenomenology.

%
\section*{ACKNOWLEDGEMENTS}
The authors would like to thank V. Bertone for their valuable comments.
F. Irani and K. Azizi are thankful to Iran National Science Foundation (INSF) for financial support provided for this research under grant No. 40406911.

%
\section*{DATA AVAILABILITY}
The data that support the findings of this article are publicly available in Ref.~\cite{MMGPDsData}.

%

%

\bibliographystyle{apsrev4-1}
\bibliography{article} 

\end{document}